\documentclass[a4paper]{jpconf}
\usepackage[square,sort&compress]{natbib}
\usepackage{graphicx}

\usepackage{amssymb}

\begin{document}
\title{Astrophysical Relevance of Storage-Ring Electron-Ion Recombination Experiments}

 \author{Stefan Schippers}

 \address{Institut f\"{u}r Atom- und Molek\"{u}lphysik, Justus-Liebig-Universit\"{a}t, 35392 Giessen, Germany}

 \ead{Stefan.Schippers@iamp.physik.uni-giessen.de}

\begin{abstract}
The relevance of storage-ring electron-ion recombination experiments for astro\-physics is outlined. In particular, the role of low-energy dielectronic-recombination resonances is discussed.
A bibliographic compilation of electron-ion recombination measurements with cosmically abundant ions is provided.
\end{abstract}

\section{Introduction}

Heavy-ion storage rings equipped with electron coolers are an excellent experimental environment for
electron-ion collision studies. Some recent studies of dielectronic recombination (DR) focussed on
high-resolution spectroscopy of highly-charged ions. Highlights of this research are the measurement of
the hyperfine induced decay rate of the $1s^2\,2s\,2p\;^3P_0$ state in berylliumlike Ti$^{18+}$ \cite{Schippers2007a}
utilizing DR at the storage ring TSR of the Heidelberg Max-Planck-Institute for Nuclear Physics, the
observation of the isotope shift in DR of three-electron Nd$^{57+}$ \cite{Brandau2008a} using different isotopes of this
ion at GSI's storage ring ESR and the observation of the hyperfine splitting of Sc$^{18+}$ low-energy  DR
resonances at the TSR high-resolution electron target \cite{Wolf2006c}. The latter experiment resulted in the
determination of the Sc$^{18+}$($2s_{1/2} - 2p_{3/2}$) energy splitting with an uncertainty of only 4.6
ppm which is less than 1\% of the few-body effects on radiative corrections \cite{Lestinsky2008a}.
Since these exciting developments have already been reviewed recently \cite{Schippers2008a}, the present review focusses on the
relevance of storage-ring electron-ion experiments for astrophysics.

Storage-ring experiments provide particularly valuable information on DR in low-temperature plasmas such as photoionized plasmas that occur, e.g., in
active galactic nuclei (AGN) in the vicinity of super-massive black holes \cite{Savin2007d}. In such plasmas highly charged ions
exist at relatively low temperatures. For many ions, the DR rate coefficients, that determine the charge
balance in these plasmas, depend sensitively on the low-energy DR resonance structure at relative electron-ion energies below $\lesssim$ 3 eV. In the following, the influence of low-energy DR resonances on electron-ion recombination rate coefficients in low-density plasmas and recent efforts of building a recombination data base for astrophysical modeling of photoionized plasmas are briefly discussed. Finally, a compilation of experimental results for the astrophysically most abundant ions is presented.

\section{Low-energy dielectronic recombination resonances}

\begin{figure}[b]
\includegraphics[width=\textwidth]{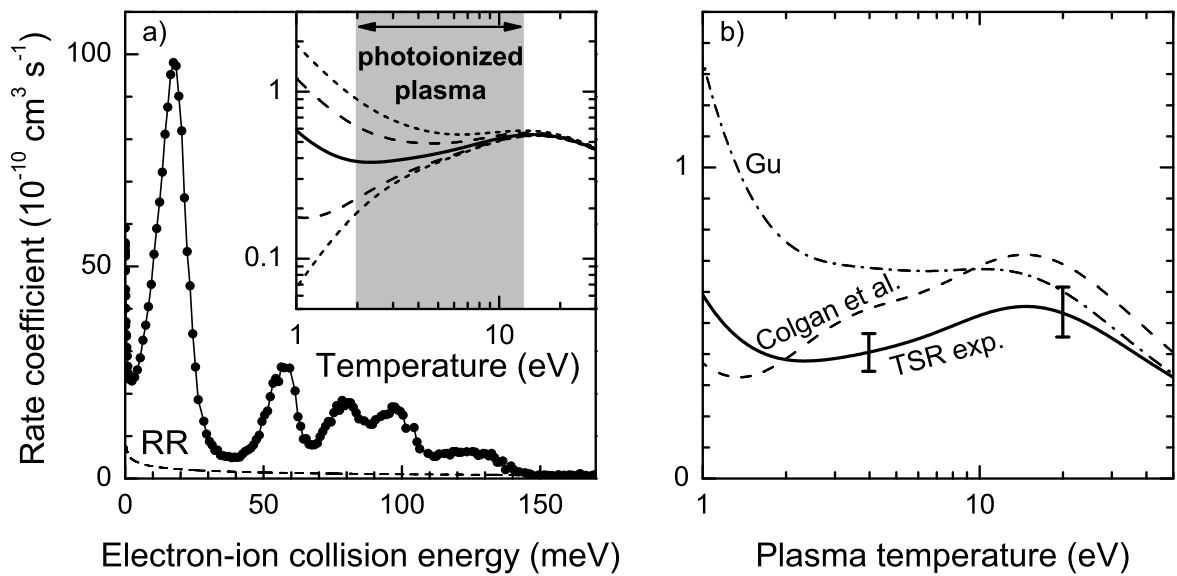}
\caption{\label{fig:Mg8}a) Experimental merged-beams rate coefficients for the recombination of electrons with Be-like Mg$^{8+}$ ions at low electron-ion collision energies measured at the Heidelberg heavy-ion storage ring TSR \cite{Schippers2004c}. The theoretically calculated contribution by non-resonant radiative recombination (RR) is shown as a dashed line. The inset shows the effect of hypothetical resonance shifts by $\pm$50 meV (dashed curves) and by $\pm$100 meV (dotted curves)
on the Mg$^{8+}$ recombination-rate coefficient in a plasma. The temperature range where Mg$^{8+}$ occurs in a photoionized plasma is highlighted. b)
The experimental Mg$^{8+}$ recombination-rate coefficient in a plasma \cite{Schippers2004c} at low temperatures compared with recent state-of-the-art theoretical results by Colgan et al.\ \cite{Colgan2003a} (dashed curve) and Gu \cite{Gu2003b} (dash-dotted curve).}
\end{figure}

\begin{figure}[b]
\includegraphics[width=\textwidth]{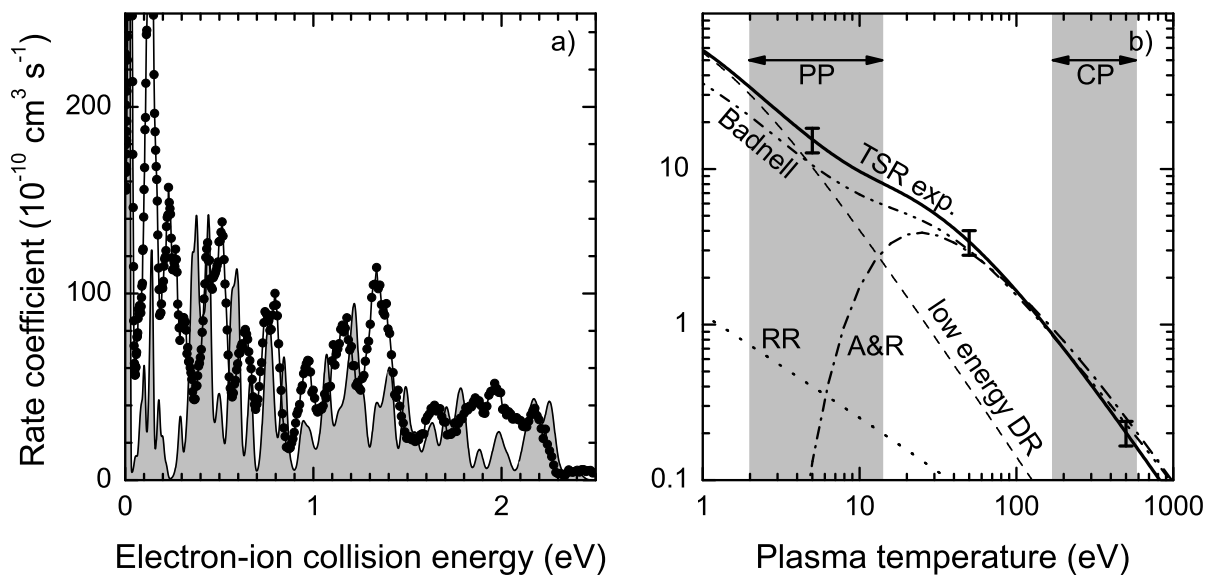}
\caption{\label{fig:Fe13}a) Experimental (filled circles) \cite{Schmidt2006a} and theoretical (shaded curve) \cite{Badnell2006b}  merged-beams rate coefficient
for the recombination of electrons with Al-like Fe$^{13+}$ ions at low electron-ion collision energies. b)  Fe$^{13+}$ recombination-rate coefficient in a plasma. The experimental result from the TSR (full curve) is compared with the theoretical rate coefficient from the widely used compilation of Arnaud \& Raymond  \cite{Arnaud1992} (dash dotted curve) and the recent state-of-the-art calculation of Badnell \cite{Badnell2006b} (dash-dot dotted curve). The contribution of RR to the latter is shown as a dotted curve. The dashed curve is the contribution of the low-energy DR resonances shown in panel a) to the total plasma rate coefficient. The temperature ranges where Fe$^{13+}$ exists in photoionized plasmas (PP) and collisionally ionized plasmas (CP) are highlighted.}
\end{figure}

Figure \ref{fig:Mg8}a shows the measured Mg$^{8+}$ recombination rate coefficient \cite{Schippers2004c} at electron-ion collision energies below 200~meV.
The strongest dielectronic recombination (DR) resonance occurs at 21~meV. The inset shows that the rate coefficient in plasma changes by more
than an order of magnitude when the low-energy resonances are shifted by as little as $\pm$100~meV. Usually, for ions with more than three electrons, the theoretical uncertainties for DR resonance energies are much larger. Correspondingly, the uncertainties of calculated recombination rate coefficients in a plasma can be rather large, especially at low temperatures where photoionized plasmas exist. As an example figure\ \ref{fig:Mg8}b compares the experimentally derived Mg$^{8+}$ rate coefficient in a plasma with two recent state-of-the-art calculations. The deviations of the theoretical
from the experimental results are as large as a factor of 2 and, additionally, there is considerable discrepancy between both theoretical curves.

This situation can be regarded as typical. Even state-of-the-art atomic codes are not always capable of providing reliable low-temperature DR plasma rate coefficients for ions. The difficulty in calculating sufficiently precise DR resonance energies is rooted in the many-body nature of the problem. Only in special cases, where one electron is outside a closed shell, relativistic perturbation theory (RMBPT) yields reliable low-energy DR resonance parameters as has been demonstrated, e.g., for Li-like fluorine \cite{Tokman2002}, sodium \cite{Nikolic2004a}, and scandium \cite{Lestinsky2008a,Wolf2006c,Kieslich2004a} as well as for Na-like silicon \cite{Orban2007a}. However, RMBPT cannot easily be extended to more complex ionic systems and, therefore, cannot satisfy the vast astrophysical atomic data needs.

\section{Dielectronic recombination data for photoionized plasmas}

Cosmic atomic plasmas can be divided into broad classes, collisionally ionized plasmas (CP) and photoionized plasmas (PP) \cite{Savin2007d}. Historically, most theoretical recombination data were calculated for CP where highly charged ions exist only at rather large temperatures, e.g., in the solar corona. At these temperatures, recombination rate coefficients are largely insensitive to low-energy DR resonances. Consequently, the theoretical uncertainties are much smaller at higher than at lower plasma temperatures which are typical for PP. Until recently, theoretical recombination rate coefficients were mainly calculated for the CP temperature ranges (see e.g. \cite{Mazzotta1998}). If these data are used for the astrophysical modeling of PP, inconsistencies arise. This has been noted, e.g., in the astrophysical modeling of x-ray spectra from AGN by Netzer \cite{Netzer2004a} and Kraemer et al. \cite{Kraemer2004a}. On the basis of astrophysical modeling these authors have suspected that the DR rate coefficients for iron ions with an open M-shell (Fe$^{1+}$--Fe$^{15+}$) from the widely used compilation of Arnaud \& Raymond \cite{Arnaud1992} are much too low in the PP temperature range.

These findings motivated storage-ring recombination measurements with iron M-shell ions. Figure \ref{fig:Fe13} shows  results, that were obtained for Al-like Fe$^{13+}$ ions. The low-energy merged-beams recombination rate coefficient (figure \ref{fig:Fe13}a) is dominated by strong DR resonances that decisively determine the low-temperature recombination rate coefficient in a PP (dashed line in figure \ref{fig:Fe13}b). As suspected, the experimentally derived PP rate coefficient is considerably higher --- by up to orders of magnitude --- than the early theoretical result from the compilation of Arnaud \& Raymond \cite{Arnaud1992} (figure \ref{fig:Fe13}b). Such discrepancies have also been found for other iron M-shell ions (see references in table\ \ref{tab:DR}).

It is clear, that the large discrepancies between the experimental and the early theoretical rate coefficients are due to a simplified theoretical treatment that was geared towards CP and more or less disregarded low-energy DR in order to keep the calculations tractable. Modern computers allow more sophisticated approaches, and recent
theoretical work has aimed at providing a more reliable recombination data-base by using state-of-the-art atomic codes \cite{Badnell2007b}.
Badnell and coworkers \cite{Badnell2003a} have calculated DR rate coefficients for finite-density plasmas. Results have been published for the isoelectronic sequences from H-like to Mg-like \cite{Altun2007a} (and references therein). Complementary RR rate coefficients are also available \cite{Badnell2006d}. Independently, Gu calculated DR and RR rate coefficients for selected ions of astrophysical interest \cite{Gu2003b,Gu2003c,Gu2004a}. Figure \ref{fig:Mg8}b shows the Mg$^{8+}$ DR results from these two data sets. Although the new theoretical work removes the striking low-temperature disagreement that was found between experimental and early theoretical results, uncertainties remain as discussed above.

For the recombination of Fe$^{13+}$ a detailed comparison between the experimental and new theoretical results was presented by Badnell \cite{Badnell2006b}.
Experimental and calculated low-energy merged-beams recombination rate coefficients are shown in figure \ref{fig:Fe13}a. Although calculated and measured DR resonance structures are approximately of the same height there are many differences in the details. Fortuitously, these differences largely average out in
the plasma recombination rate coefficient. Still, the theoretical plasma rate coefficient is up to 50\% smaller than the experimental one in the PP temperature range (figure \ref{fig:Fe13}b).

In conclusion, storage ring experiments are presently the only reliable source for low-temperature DR data. Moreover, they provide valuable benchmarks for the further development of the theoretical methods.

\section{Bibliography of experimental results from storage rings}

\begin{figure}[b]
\includegraphics[width=0.512\textwidth]{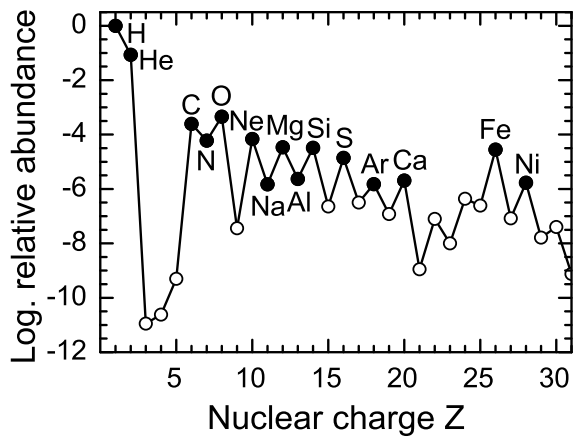}
\begin{minipage}[b]{0.45\textwidth}\caption{\label{fig:relabundance}Abundances of the elements in the solar photosphere relative to hydrogen
\cite{Asplund2006a}. The astrophysically most relevant elements are marked by solid circles. Elements with nuclear charge $Z>30$ have relative abundances below $10^{-8}$. Iron is the heaviest element that is produced by nuclear fusion in stars. Iron ions play a prominent role in x-ray astronomy \cite{Kahn2002a} due to their large relative abundance and their high nuclear charge.}
\end{minipage}
\end{figure}

For the interpretation and understanding of astronomical observations numerical models of astrophysical plasmas are used \cite{Ferland2003a}. As an input these models require atomic data especially for the 15 most abundant elements \cite{Kallman2007a}. Their relative abundances in the solar photosphere \cite{Asplund2006a} are depicted in figure \ref{fig:relabundance}. Table \ref{tab:DR} lists those astrophysically relevant ions for which experimental results on electron-ion recombination have been obtained from storage-ring measurements. The corresponding references are given. More comprehensive but now somewhat outdated compilations which additionally contain results from early single-pass merged-beams experiments as well as results for other ions have been published previously \cite{Mueller1995,Schippers1999d,Schippers2002c}.

\begin{table}[t]
 \caption{\label{tab:DR}Bibliographic compilation of storage-ring electron-ion recombination experiments with astrophysically relevant ions.  For each ion the primary charge state before recombination is listed.}
 \begin{center}
\begin{tabular}[c]{l@{\,}rc@{~~~~~~~~~~}l@{\,}rc@{~~~~~~~~}l@{\,}rc}
 \br
 \multicolumn{2}{c}{ion}  & reference& \multicolumn{2}{c}{ion}  & reference & \multicolumn{2}{c}{ion}  & reference\\
 \mr
 H  & 1+     & \cite{Biedermann1995,Gao1997}                    &Ne & 6+     & \cite{Orban2008a}                         &Fe  & 7+     & \cite{Schmidt2007a,Schmidt2008a}            \\
 He & 1+     & \cite{Tanabe1992,Haar1993,Dewitt1994,Dewitt1995} &   & 7+     & \cite{Zong1998,Boehm2001b,Boehm2005a}     &  & 8+     & \cite{Schmidt2007a,Schmidt2008a}              \\
    & 2+     & \cite{Gao1997}                                   &Ne & 10+    & \cite{Gao1995,Gao1997}                    &  & 9+     & \cite{Lestinsky2008b}            \\
 C  & 2+     & \cite{Fogle2005a}                                &Na & 8+     & \cite{Nikolic2004a}                       &  & 10+    & \cite{Lestinsky2008b}            \\
    & 3+     & \cite{Mannervik1998,Schippers2001c}              &Mg & 8+     & \cite{Schippers2004c}                     &  & 13+    & \cite{Schmidt2006a}  \\
    & 4+     & \cite{Kilgus1993,Mannervik1997}                  &Si & 3+     & \cite{Orban2006a,Orban2007a,Schmidt2007b} &  & 14+    & \citep{Lukic2007a}   \\
    & 5+     & \cite{Wolf1991}                                  &   & 11+    & \cite{Kenntner1995,Bartsch1997}           &  & 15+    & \cite{Linkemann1995c}\\
    & 6+     & \cite{Gwinner2000}                               &   & 14+    & \cite{Gao1997}                            &  & 16+    & \cite{Schmidt2008b}       \\
 N  & 3+     & \cite{Fogle2005a}                                &S  &  5+    & \cite{Orban2008b}                         &  & 17+    & \cite{Savin1997,Savin1999}       \\
    & 4+     & \cite{Glans2001,Boehm2005a}                      &   & 15+    & \cite{Wolf1992b}                          &  & 18+    & \cite{Savin1999,Savin2002c}      \\
    & 7+     & \cite{Gao1997}                                   &Ar &  7+    & \cite{Orban2008b}                         &  & 19+    & \cite{Savin2002a} \\
 O  & 4+     & \cite{Fogle2005a}                                &   & 13+    & \cite{Gao1995,Dewitt1996}                 &  & 20+    & \cite{Savin2003a} \\
    & 5+     & \cite{Boehm2002a,Boehm2003a}                     &   & 15+    & \cite{Zong1997}                           &  & 21+    & \cite{Savin2003a}\\
    & 7+     & \cite{Kilgus1990}                                &   &        &                                           &  & 22+    & \cite{Savin2006a}\\
    &        &                                                  &   &        &                                           &Ni& 17+    & \cite{Fogle2003a,Fogle2003b}  \\
    &        &                                                  &   &        &                                           &  & 25+    & \cite{Schippers2000b}\\  \br
 \end{tabular}
 \end{center}
 \end{table}

\ack

The author is grateful to Michael Lestinsky, Alfred M\"{u}ller, Daniel Wolf Savin, Eike Schmidt and Andreas Wolf  for long-standing fruitful collaboration
and likewise to all his other collaborators who are too many to list them all.
This research was supported in part by the German federal research-funding
agency DFG under contract no.\ Schi~378/5.

\medskip


\providecommand{\newblock}{}

\end{document}